\documentclass[aps,prl,twocolumn,showpacs]{revtex4}

\newcommand{\lc}{\xi_{\rm c}}

\newcommand{\lH}{\xi_H}

\newcommand{\Tg}{T_{\rm g}}

\newcommand{\Tc}{T_{\rm c}}

\newcommand{\Meq}{M_{\rm eq}}
\newcommand{\chieq}{\chi_{\rm eq}}

\newcommand \be {\begin{equation}}
\newcommand \ee {\end{equation}}
\usepackage{graphicx}
\usepackage{epsf}
\usepackage{amssymb}
\usepackage{array}
\usepackage{color}
 
\begin{document}
 
\title{Dynamical breakdown of the Ising spin-glass order under a magnetic field}
 
\author{
P. E. J{\"o}nsson\cite{newadd}
}
\affiliation{ISSP, University of Tokyo, Kashiwa-no-ha 5-1-5, Kashiwa, Chiba
277-8581, Japan
}
 
\author{
H. Takayama
}
\affiliation{ISSP, University of Tokyo, Kashiwa-no-ha 5-1-5, Kashiwa, Chiba
277-8581, Japan
}
 
\author{H. Aruga Katori}
\affiliation{
RIKEN, Hirosawa 2-1, Wako, Saitama, 351-0198, Japan
}
\author{A. Ito}
\affiliation{
RIKEN, Hirosawa 2-1, Wako, Saitama, 351-0198, Japan
}

\date{\today}
 
\begin{abstract}
 The dynamical magnetic properties of an Ising spin glass
 Fe$_{0.55}$Mn$_{0.45}$TiO$_3$ are studied under various magnetic
 fields. Having determined the temperature and static field 
 dependent relaxation time $\tau(T;H)$ from ac magnetization
 measurements under a dc bias field by a  general method, we first
 demonstrate that these data provide evidence for a spin-glass (SG)
 phase transition only in zero field. We next argue that the data
 $\tau(T;H)$ of finite $H$ can be well interpreted by the droplet theory 
 which predicts the absence of a SG  phase transition in finite fields.
\end{abstract}
                                                                            

\maketitle
 
One of the recent interests in the spin-glass (SG) study is the nature  
of Heisenberg spin glasses under a magnetic field~\cite{Petit,imakaw2004}.
There exists, however, a still unsettled problem even on the
conceptually much simpler Ising spin glasses. The mean-field theory
predicts that the SG phase remains up to the  de
Almeida-Thouless (AT) line in the $(H,T)$ plane~\cite{almtho78}. 
The droplet theory, based on the short-range Edwards-Anderson (EA)
model, instead predicts that any applied magnetic field will break up
the equilibrium SG long-range order~\cite{fishus88eq,fishus88noneq}.
Experimental investigations on Ising spin glasses give evidence both for
the existence of an AT-line~\cite{aruito94} and against critical
dynamics under a field~\cite{matetal95}, while some recent simulation
and experimental results indicate that the SG phase does not exist under
a magnetic field~\cite{takhuk2004,youkat2004,krz,PEJ-HT}.
 
In the present paper, we address the problem through ac susceptibility
measurements on the Ising spin glass Fe$_{0.55}$Mn$_{0.45}$TiO$_3$ under
a dc bias field $H$ at temperature $T$. We first specify, in a manner
explained in detail below, sets of $(T,H)$, at which the characteristic
relaxation time of the system, $\tau(T;H)$, coincides with the inverse
of the ac field frequency $\omega$, i.e., $1/\omega=\tau(T;H)$. We then
examine whether or not these $\tau(T;H)$ obey a dynamical critical
behavior represented by
\be
 \tau(T;H) \sim t_0(\lc/L_0)^z \sim t_0|T/\Tg(H)-1|^{-z\nu},
\label{eq-tc}
\ee 
at $T>\Tg(H)$. Here we suppose that the SG replica symmetry breaking
(RSB) phase predicted to appear below the AT line, $\Tg(H)$, by the
mean-field theory is accompanied with a high-temperature disordered
phase which exhibits critical divergences of the correlation length
$\xi_c(T;H) \sim L_0 |T/\Tg(H)-1|^{-\nu}$ and of the correlation time
given by the above equation, where $t_0$ and $L_0$ are respectively
microscopic units of time and length, and $z$ and $\nu$ are critical
exponents. It turns out that the data $\tau(T;H)$ are not compatible
with the expected dynamical critical behavior except for the case with
$H=0$.   

In the droplet theory, on the other hand, the so-called magnetic
correlation length, $\lH$, is introduced. It specifies the behavior of
droplet excitations with a linear size $L$ in a SG state under a field
$H$~\cite{fishus88eq}. For droplets with $L < \lH$ their behavior is
governed by the SG stiffness free energy $\Upsilon(T)(L/L_0)^\theta$,
while for those with $L > \lH$ their behavior is dominated by the Zeeman
energy $\sqrt{q_{\rm EA}} H (L/L_0)^{d/2}$, where $\Upsilon(T)$ is the
SG stiffness modulus, $\theta$ the stiffness exponent, $q_{\rm EA}(T)$
the EA order parameter, and  $d$ the spatial dimension. 
Explicitly, at $T$ less than $\Tc [=T_g(0)]$ which is a 
unique critical point of the system, $\lH$ is written as 
\be
\lH \sim L_0 \left(\frac{\Upsilon(T)}{H\sqrt{q_{\rm EA}(T)}}\right)^\delta
\sim L_0 \left(\frac{(1-T/\Tc)^{a_{\rm eff}}}{H}\right)^\delta,
\label{LH}
\ee
where $\delta=2/(d-2\theta)$. In the last expression the temperature
dependence of $\Upsilon/\sqrt{q_{\rm EA}}$ is represented by 
$(1-T/\Tc)^{a_{\rm eff}}$. At $T \lesssim \Tc$, 
$a_{\rm eff}=\theta \nu-\beta/2$ is expected, where $\beta$ is the
critical exponent of $q_{\rm EA}$. In the present analysis on the ac
susceptibility measurements of frequency $\omega$ under $H$, we
identify $\lH$ with $L_T(t=1/\omega)$, which is the mean size of
droplets that can respond to the ac field of frequency $\omega$ at
temperature $T$, i.e., 
\be
\lH \approx L_T(1/\omega).
\label{LH-LT}
\ee
Furthermore it is considered that $L_T(t)$ has the same functional
form as that of the growth law $R_T(t)$ of the SG correlation length
which grows with time $t$ after the system is kept at a constant
temperature $T$ under $H=0$. Explicitly, we adopt here an algebraic
growth law 
\be
L_T(t) \sim R_T(t) \sim L_0(t/t_0)^{b T/\Tc},
\label{eq-pow}
\ee
which is commonly observed in numerical
simulations~\cite{kisetal96,maretal98,komyostak99} as well as in
experiments~\cite{beretal2001,jonetal2002PRL}. Interestingly,
the present data $\tau(T;H)$ for relatively large $H$ and so relatively
low $T$ turn out to be consistent with the scenario represented by
Eqs.~(\ref{LH}), (\ref{LH-LT}), and (\ref{eq-pow}). We interpret this
result as evidence for the presence of the magnetic correlation length
$\lH$ predicted by the droplet theory.  

\begin{figure}[bht]
\includegraphics[width=0.75\columnwidth]{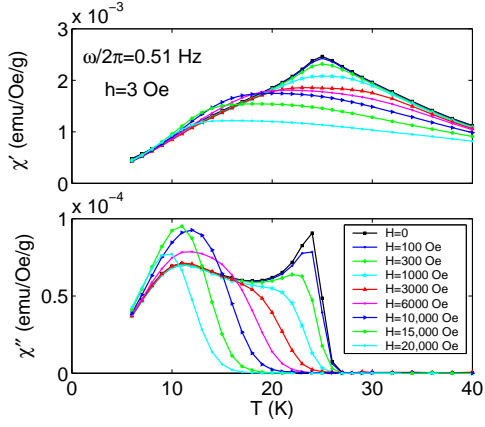}
\caption{(Color online) ac susceptibility vs temperature for bias fields $H=0$, 100, 300, 1000, 3000, 6000, 10\,000, 15\,000, and 20\,000~Oe. }
\label{ac}
\end{figure}

\begin{figure}[bht]
\includegraphics[width=0.75\columnwidth]{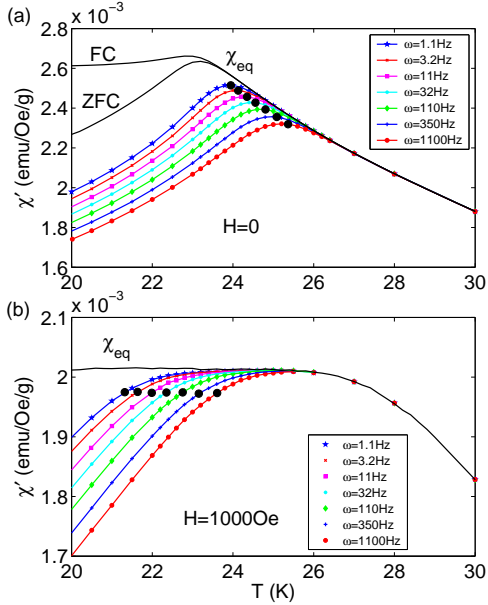}
\vspace*{-1mm}
\caption{(color online) (a) $M_{\rm ZFC}/h$,  $M_{\rm FC}/h$ and $\chi'(T,H;\omega$) vs temperature in zero bias field. (b) $\chieq$ and $\chi'$ vs temperature in a bias field of $H=1000$~Oe.
Dots indicate the condition of Eq.~(\ref{condition}).}
\label{acH0}
\end{figure}

We measure the ac susceptibility, 
$\chi(\omega,T,H)=\chi'(\omega,T,H)+i \chi''(\omega,T,H)$, of the Ising
spin glass Fe$_{0.55}$Mn$_{0.45}$TiO$_3$ in the frequency range of 
$0.001~{\rm s} \lesssim 1/\omega \lesssim 1~{\rm s}$.  All
measurements are performed on a MPMS-XL squid magnetometer
equipped with the low-field option. The susceptibilities $\chi'$
and $\chi''$ measured under different bias fields are shown in
Fig.~\ref{ac}, while shown in Fig.~\ref{acH0} are the dc and ac
susceptibilities of different frequencies in zero bias field and in a
bias field of $H=1000$~Oe. The equilibrium susceptibility $\chieq(T,H)$
can be determined in the temperature range within which the dc
field-cooled (FC) and zero-field-cooled (ZFC) magnetization $M$ 
(measured with a slow cooling rate) coincide with each other, as
$
\chieq(T,H)=\frac{d\Meq(T,H)}{dH} \approx
\frac{\Delta \Meq(T,H)}{\Delta H} 
=\frac{\Meq(T,H+h)-\Meq(T, H-h)}{2h}
$.
The field $h$ should be small in order to probe the linear response.

\begin{figure}[bht]
\includegraphics[width=0.75\columnwidth]{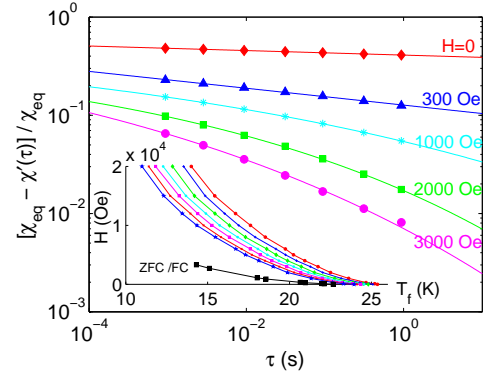}
\caption{(Color online) $[\chi_{\rm eq}-\chi'(\tau)]/\chi_{\rm eq}$ vs time on a double 
 logarithmic scale. The lines are fits to a power law for  $H=$0 and
 300~Oe or to a stretched-exponential for $H=$1000, 2000 and
 3000~Oe. Inset: the sets of $(T,H)$ determined by
 Eq.~(\ref{condition}) for  $\tau(T;H)$ equal to $9.4\cdot10^{-1},\,3.1\cdot10^{-1},\,9.4\cdot10^{-2},\, 3.1\cdot10^{-2},\,9.4\cdot10^{-3},\, 2.8\cdot10^{-3},\,9.4\cdot10^{-4}$~s (left to right).
}
\label{TfH_q}
\end{figure}

As indicated by circles in Fig.~\ref{acH0}, we identify the 
characteristic relaxation time $\tau(T;H)\ (=1/\omega)$ by the
condition~\cite{HJNS}, 
\be
\chi'(\omega,T,H) =0.98\chieq(T,H).
\label{condition}
\ee
To explain our idea behind this condition we present in Fig.~\ref{TfH_q}
$\chi_{\rm eq}-\chi'(\omega)$, which is proportional to the dynamic spin
correlation function $q(t)$ with $t=1/\omega$, for different bias fields
at $T=20$~K. It has been found to follow the empirical formula  
$q(t)=Ct^{-\alpha}e^{-(t/\tau^*(T))^y}$ at $T$ above $\Tc$ and a pure
power law below $\Tc$ for $H=0$ both in numerical simulations on the EA
Ising model\cite{ogielski} and in experiments on 
Fe$_{0.5}$Mn$_{0.5}$TiO$_3$\cite{gunnarsson}. The results are
interpreted to indicate that the distribution of relaxation times is
bounded from above by a certain value around $\tau^*(T)$ at $T>\Tc$,
while $\tau^*(T)$ is 
        infinite at $T \leq \Tc$. Although our experimental
timewindow for a fixed set of $(T,H)$ is rather limited,
Fig.~\ref{TfH_q} indicates that it is possible to bring the Ising SG to
equilibrium by applying a magnetic field. In particular, for the set of
$(T,H)$ for which Eq.~(\ref{condition}) is found to be satisfied, the
corresponding $q(t)$ exhibits a stretched-exponential form and so the
corresponding $\tau^*(T;H)$ is definitely finite. We have therefore
simply introduced the condition of Eq. (\ref{condition}) to specify the
upper bound of relaxation times without fitting our $q(t)$ to a
stretched-exponential form explicitly. We consider that the present
method for specifying $\tau(T;H)$ is systematic and preferable to 
other methods used previously\cite{foot1}. It is also noted that the
condition is satisfied in the temperature range where we can determine
$\chi_{\rm eq}$ as seen in Fig.~\ref{acH0}, implying that $\tau(T;H)$
thus specified in fact associates with certain relaxation processes in
equilibrium.    

\begin{figure}[bht]
\includegraphics[width=0.75\columnwidth]{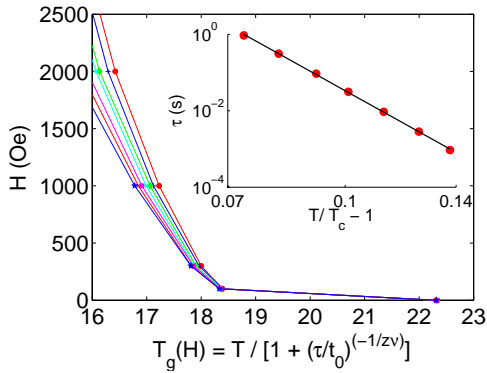}
\caption{(Color online) Test for critical dynamics assuming that the spin glass
 transition temperature 
 $\Tg(H)=T(\tau;H)/[1+(\tau/t_0)^{-1/z\nu}]$
 [Eq.~(\ref{eq-tc})] changes with $\tau$ for each $H$ as explained in
 the text.
 The inset shows 
    $\tau$ vs $T(\tau;H=0)/\Tc-1$ with
 $\Tc=\Tg(H=0)=22.3$~K on a log-log scale. 
}
\label{TgH}
\end{figure}

In the inset of Fig.~\ref{TfH_q}, we show the $H-T$ relations which
yield common values of $\tau(T;H)$. They are roughly consistent with the
AT-lines, i.e., of the form $H \propto [1-T(\tau;H)/T(\tau;0)]^{3/2}$. 
However, such apparent AT-lines dependent on measuring time scales
$\tau$ are by no means a proof of an equilibrium SG phase under a
field~\cite{wenmyd84}. We require the dynamical critical slowing down
represented by Eq.~(\ref{eq-tc}) to hold at $T>\Tg(H)$ where $\Tg(H)$ is
an assumed critical temperature under a field $H$. As shown in the inset
of Fig.~\ref{TgH}, $\tau(T;0)$ in our timewindow are in fact fitted to
this expression with $\Tc=\Tg(0) \approx 22.3$~K, 
$t_0\approx 3\cdot 10^{-13}$~s, and $z\nu \approx 11$. The value of
$z\nu$ is in accordance with the previous values obtained for the Ising
spin glass Fe$_{0.50}$Mn$_{0.50}$TiO$_3$~\cite{matetal95}. This result
gives evidence for an equilibrium  SG phase transition in zero field.  
The main frame of Fig.~\ref{TgH} is, on the other hand, the result of an
attempt to see if the system also exhibits critical slowing down in
small dc fields assuming that $t_0$ remains the same as in zero field. 
The best fit to Eq.~(\ref{eq-tc}) under this constraint is found with
$z\nu \approx 22$ for all $H>0$. To demonstrate the quality of the fit
we show a $H-T$ diagram whose $T$-axis is $\Tg(H)$ calculated for each 
$\tau(T;H)$ using the obtained value of $z\nu$. A unique $\Tg(H)$ is
found only for $H=100$~Oe (and $H=0$) and the data in $H>100$~Oe exhibit
systematic dispersion, giving evidence against critical dynamics for
$H>100$~Oe. Even for $H=100$~Oe, however, due to the large value of
$z\nu \approx 22$ we consider that the relatively good fit to
Eq.~(\ref{eq-tc}) has no physical meaning. We also note that if both
$t_0$ and $z\nu$ are adjusted in the fitting, the resultant $t_0$ takes
unphysically large values. We therefore conclude that, under a magnetic
field, the system does not exhibit a phase transition which is
accompanied with critical slowing down. 
 
\begin{figure}[thb]
\includegraphics[width=0.75\columnwidth]{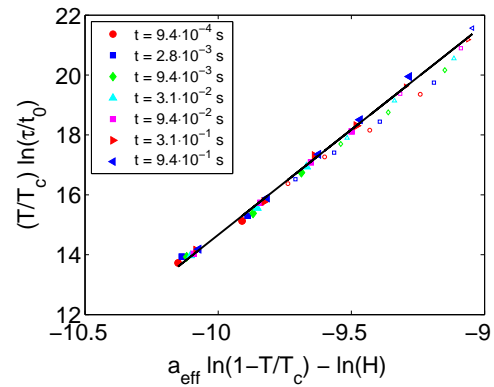}
\caption{(Color online) The scaling plots of $(T/T_c)\ln\tau(T;H)$ vs $a_{\rm eff}{\rm ln}(1-T/\Tc) - {\rm ln}H$ with $a_{\rm eff}=0.25$.
Open (filled) symbols mark temperatures $\geq$ ($<$) $0.7\Tc$.
}
\label{LHcorr}
\end{figure}

Next, let us examine the data $\tau(T;H)$ for relatively large $H$
based on the droplet picture, namely, by regarding $\tau(T;H)$ as a   
relaxation time of the largest SG clusters of a mean size $\lH(T;H)$,
which is determined by Eq.~(\ref{LH-LT}) combined with Eqs.~(\ref{LH})
and (\ref{eq-pow}). For this purpose we plot $(T/\Tc)\ln(\tau/t_0)$
for $H \ge 5000$~Oe as a function of $a_{\rm eff}\ln(1-T/\Tc) - \ln H$,
thereby adjusting $a_{\rm eff}$ but keeping $t_0$ fixed to the value
obtained above. As shown in Fig.~\ref{LHcorr}, the best collapse of the
data at $T/\Tc \lesssim 0.7$ is obtained with 
$a_{\rm eff} \approx 0.25$. The slope of the fit, which is equal to
$b/\delta$, gives $b \approx 0.11$, where $\theta=0.2$~\cite{bramoo84}
(and so  $\delta=$0.77) are used. Here we examine $\tau(t;H)$ 
in a
rather narrow temperature range (the lower bound of our observation of
$\tau$ is $0.55\Tc$), 
 since the approximation 
$\Upsilon/{\sqrt q_{\rm EA}}\sim (1-T/\Tc)^{a_{\rm eff}}$ used to derive
the last expression of Eq.~(\ref{LH}) with a constant $a_{\rm eff}$ is
not expected to work in a wider temperature range. In fact when we
analyze the $\tau(T;H)$ data obtained for 
Fe$_{0.5}$Mn$_{0.5}$TiO$_3$ by Mattsson {\it et al.}~\cite{matetal95} by
the present method, we obtain $a_{\rm eff}\approx 0$ and 
$b \approx 0.09$ for  their data at $T \lesssim 0.5\Tc$, and 
$a_{\rm eff} \approx 0.2$ and $b \approx$ 0.09 for their data at 
$0.5\Tc \lesssim T \lesssim 0.7\Tc$. The circumstances become more
subtle at $T$ closer to $\Tc$, since the critical behavior of
$\Upsilon(T)$ and 
    $q_{\rm EA}(T)$ 
has not been well established yet. We can only mention
that the same analysis on our data $\tau(T;H)$ at $T \lesssim 0.95\Tc$
and for $H\lesssim 5000$~Oe yields $a_{\rm eff} \approx 0.45$ and 
$b \approx 0.13$. This strongly implies that the critical exponent, 
$a_{\rm eff}$ at $T=\Tc$, is positive. In spite of such subtlety
concerning with the temperature dependences of $\Upsilon(T)$ and $q_{\rm
EA}(T)$, the results shown in Fig.~\ref{LHcorr} are sufficient for us to
conclude that the system is in the paramagnetic phase,  with 
$\lH$ and $\tau(T;H)$ being the upper bounds for the  SG correlation length and relaxation time, respectively.

The values of the exponents extracted above from
$\tau(T;H)$ at $0.55\Tc \lesssim T \lesssim 0.7\Tc$ can be compared
with those obtained in the simulation on the Ising EA model in the
corresponding temperature range; $b \approx 0.16$~\cite{komyostak99}
and $a_{\rm eff} \approx 0.14$, where the latter value is extracted from
$\lH$ which is obtained and denoted as $l_TH^{-\delta}$ 
in~\cite{takhuk2004}. These figures are in turn compatible with the ZFC
magnetization of Fe$_{0.5}$Mn$_{0.5}$TiO$_3$ observed in heating
processes with intermittent stops~\cite{beretal2001} as well as with the
crossover scenario~\cite{takhuk2004} for the AT-like
transition observed also in
Fe$_{0.5}$Mn$_{0.5}$TiO$_3$~\cite{aruito94}. 
Here we emphasize that the values of $b$ extracted
from the simulation and the experiments, over more than 10 decades
difference in time scales in units of $t_0$, agree with each other
reasonably well.

One more comment is in order on the growth law of the SG
correlation at a constant $T$.
In the original droplet theory~\cite{fishus88noneq}, instead of 
Eq.~(\ref{eq-pow}), a logarithmic form of 
\be
L_T(t)\sim L_0  \left[\frac{T \ln (t/t_0)}{\Delta}\right]^{1/\psi},
\label{eq-log}
\ee
has been proposed, where $\Delta$ is the characteristic energy scale of
the free-energy barrier. For Fe$_{0.50}$Mn$_{0.50}$TiO$_3$
Eq.~(\ref{eq-log}) has been applied to its various phenomena with the
resultant exponent $\psi$ ranging from 0.03 to 
1.9~\cite{matetal95,dupetal2001,jonetal2002PRL,beretal2004}. 
When Eq.~(\ref{eq-pow}) is replaced by Eq.~(\ref{eq-log})
with a $T$-independent $\Delta$ in the present analysis, we obtain
$\psi \approx 0.6$ and the same $a_{\rm eff}$ as obtained above from the
$\tau(T;H)$ data at $0.55\Tc \lesssim T \lesssim 0.7\Tc$. We consider,
however, that the power-law growth is superior to the logarithmic growth
in the sense that the values of $b$ extracted from various phenomena are
less dispersive than those of $\psi$~\cite{Bert_criticism}. The
power-law growth implies that the free-energy barrier of droplets
excitations of a size $L$ is proportional to $\ln L$ which becomes
smaller than $L^{\theta}$, the free-energy gap of the  corresponding
droplet excitations, for a sufficiently large $L$, say $L^*$. Therefore, 
our conclusion that the power-law is a more plausible description of 
various SG glassy dynamics mentioned above implies that 
the SG short-range of less than $L^*$ is
involved in such slow-dynamical phenomena observed even by laboratory
experiments. In particular, $\chi_{\rm eq}$ and $\chi'(\omega)$
analyzed in the present work involve length scales $\xi_{\rm c}(T)$ and 
$L_T(1/\omega)$ or $\lH$ which are much shorter than $L^*$ so that they can 
be regarded as SG equilibrium properties (see \cite{PEJ-HT} for 
non-equilibrium phenomena associated with similar length scales much
less than $L^*$). We believe that, in the field of glassy dynamics,
proper understanding of the length and time scales of phenomena we
observe is very important, though it is often not an easy task.

To summarize, we have found experimental evidence against the existence
of an equilibrium phase transition in Ising spin glasses under a bias
magnetic field, i.e. evidence against the AT phase transition. From the
present analysis we learn that one has also to be careful to draw
conclusions about the equilibrium phase diagram of Heisenberg spin glass
under a field; the occurrence of the irreversibility~\cite{Petit} alone
may not be evidence for the presence of an equilibrium phase transition.

We thank Per Nordblad for helpful discussions. P.E.J. acknowledges
financial support from the Japan Society for the Promotion of Science. 
The present work is supported 
by the NAREGI Nanoscience Project, from the
Ministry of Education, Culture, Sports, Science, and Technology.


\begin{thebibliography}{10}
\bibitem{foot1}
Our method here has the advantage that $\tau(T;H)$ is
determined by a common way in both measurements with and without a bias
field. The method described in Ref.~\cite{matetal95} which only involves
measurements of the ac susceptibility without a bias field requires that
$\chi(\omega,T,H)=\chi(\omega,T,H=0)$ at $T$ of interest, which is not
always the case for the present sample as can be seen in Fig.~\ref{ac}.
\end{thebibliography}
\end{document}